# Making drawings speak through mathematical metrics

Cédric Sueur[1,2], Lison Martinet[1], Benjamin Beltzung[1], Marie Pelé[3]

1: Université de Strasbourg, CNRS, IPHC UMR 7178, Strasbourg, France

2: Institut Universitaire de France, Paris, France

3 : Anthropo-Lab, ETHICS EA7446, Lille Catholic University, Lille, France

Corresponding author: Cédric Sueur, cedric.sueur@iphc.cnrs.fr; IPHC UMR 7178, 23 rue Becquerel, 67087 Strasbourg, France

**Abstract:**

Figurative drawing is a skill that takes time to learn, and evolves during different childhood phases that begin with scribbling and end with representational drawing. Between these phases, it is difficult to assess when and how children demonstrate intentions and representativeness in their drawings. The marks produced are increasingly goal-oriented and efficient as the child's skills progress from scribbles to figurative drawings. Pre-figurative activities provide an opportunity to focus on drawing processes. We applied fourteen metrics to two different datasets (N=65 and N=345) to better understand the intentional and representational processes behind drawing, and combined these metrics using principal component analysis (PCA) in different biologically significant dimensions. Three dimensions were identified: efficiency based on spatial metrics, diversity with colour metrics, and temporal sequentiality. The metrics at play in each dimension are similar for both datasets, and PCA explains 77% of the variance in both datasets. These analyses differentiate scribbles by children from those drawn by adults. The three dimensions highlighted by this study provide a better understanding of the emergence of intentions and representativeness in drawings. We have already discussed the perspectives of such findings in Comparative Psychology and Evolutionary Anthropology.

# 1. Introduction

Humans are the only species who naturally draw and paint objects. This behaviour takes time to develop over a series of childhood phases. Researchers agree that drawing evolves from scribbling in toddlers (Freeman, 1993; Kellogg, 1969) to representational drawing in older children. In its initial phases, scribbling is often viewed purely as a motor pleasure that is not guided by visual planning, and is mainly determined by the mechanical functioning of the motor system of the arm, wrist and hand (Kellogg, 1969; Luquet, 1927). The youngest children show only transient interest in their own scribbles, and often readily move from one scribble to the next (Golomb, 2003; Thomas & Silk, 1990). With increasing perceptual motor coordination and a progressive acquisition of complex and effective drawing rules, the scribbles become complex patterns that are guided by visual attention and are determined by aesthetic considerations such as balance or symmetry (Golomb, 2003; Piaget & Inhelder, 2008; Willats, 2005). While a full-blown representational drawing (i.e. preplanned by the child and readable by an observer) first appears by the age of three to four years old (Freeman, 1993; Gardner, 1981; Golomb, 2003; Piaget & Inhelder, 2008), preliminary indications of drawing-related symbolic actions can be traced back to as early as the second or third year of life. This finding suggests that at this age, children have already become aware of the dual function of a drawing, i.e. a graphic signifier that signifies a referent and is also a real object in its own right (DeLoache, 1991). Concerning the development of early mark-making and pre-representational activities, different theories have been developed, some of which are very similar (Costall, 1995; S. Cox, 2005; Matthews, 1984, 1999). The three following types of early pre-representations have been described in the literature, all of which were claimed to have appeared before children produce planned shapes: action representation, romancing, and guided elicitation (Adi-Japha et al., 1998; Matthews, 1984).

In humans, action representations, also called gestural drawings, appear both in spontaneous drawing and in response to the request to draw an object (e.g. an airplane). Children may accompany scribbling with verbalizations or sounds such as roaring, which indicates that both their motions and the marks emerging from their drawing instrument simulate the motion of an object. We can then speculate whether children are really scribbling or if they are simply demonstrating an active rather than a figurative mode of representation (Matthews, 1984). The personal actions of children are therefore not random but are intentional and combined with marks and sounds to represent the moving, roaring object (M. V. Cox, 2013; Gardner & Wolf, 1987; D. Wolf, 1988; D. Wolf & Perry, 1988). So the drawing has representativeness for the drawer (i.e. internal representativeness) even if this is not the case for an external observer (i.e. external representativeness, (Martinet et al., 2021). Romancing refers to instances where children name a scribble with an object but an observer has difficulty finding a graphic resemblance between the scribble and the object the child claims to have drawn. Naming takes place either spontaneously or when elicited by questioning from an adult, and can occur before or during the drawing, or after its completion. Action representations and romancing cannot be observed in children before they are two to three years old, or in children with certain psychopathologies. During these two stages, it is therefore difficult to establish if a drawing is goal-directed, with a meaning and an intentional representation. Evidence of intention is provided during during the third step, called guided elicitation, when the child shows no representational intention in his free drawings but produces figurative drawings when assisted. However, even if it is not always easy to demonstrate intentional pre-representational activities, we can predict without difficulty that the marks produced are increasingly goal-oriented and efficient as the child progresses from scribbles to figurative drawings.

Rather than studying the finished drawings, we should focus on the presence of pre-representative activities on drawing processes (Costall, 1995; S. Cox, 2005; Martinet & Pelé, 2020; Matthews, 1999).

Different methods have been used to answer this question of representativeness and goal directedness beyond drawing (Desmet et al., 2021; Urban, 2004). Publications describe topics ranging from the kinematic aspects of scribbling to the precursors of graphic representation, including authors who compare curved lines in drawings to mathematical laws. Children tended to attribute representational meanings (e.g. an airplane) to angular curves and nonrepresentational meanings (e.g. a line) to smooth curves that they had just finished drawing (Kellogg, 1969). However, these studies are not objective given that authors asked children *a posteriori* what their drawings represented. Moreover, this methodology cannot be applied to subjects (human or nonhuman primates) who are unable to explain their drawings. Thus, only one study to date has used methodology permitting the comparison of drawing abilities in humans and nonhuman primates or made it possible to understand the evolution of drawing in children and in other primates (Martinet et al., 2021).

This paper describes the use of different mathematical metrics to objectively and quantitatively measure intention and representativeness in drawing. Recently developed techniques make it possible to consider ethology as a physical science and apply quantitative measures in this discipline (Brown & De Bivort, 2018). Although simple drawing measures such as the number of colours can be used (Zeller, 2007), they provide few cues about the intention behind the drawing. We aim to go further by using mathematical measures to fulfil this goal. Metrics are already used to understand whether movements are optimal or evaluate the extent to which behavioural sequences are complex and predictable in animals, including humans. For instance, Martinet et al. (2021) and (Beltzung et al., 2021) used spatial and temporal fractal analyses, which had previously been applied to understand optimal movements and optimal behavioural sequences of animals searching for food (Bartumeus et al., 2002; Meyer et al., 2017, 2020; Reynolds, 2008), and found an increase in complexity and efficiency in humans compared to chimpanzees, but also an increase with age in humans. Other metrics such as entropy (Ebeling et al., 2002; Kershenbaum, 2014) or the Gini index (Debache et al., 2019; Planckaert et al., 2019) were also used to understand the distribution or complexity of different behaviours (e.g. activity, food exchange, vocalisation) from ants to humans. We used a total of fourteen metrics to enrich our understanding of the intentional and representational processes behind drawing. These metrics are detailed in Table 1 along with definitions and predictions.

This paper seeks to combine all of these metrics in different dimensions using principal component analysis. PCA is used to extract and visualize important information contained in a multivariate data table by combining metrics to form a biologically significant dimension, as already shown for personality (Bousquet et al., 2015; M. Wolf & Weissing, 2012) or sociality (Viblanc et al., 2016). Here, we expect metrics to be combined and form dimensions that correspond to representativeness (at least internal, meaning from the point of view of the drawer but not for the observer) and show evidence of anticipation or aestheticism in the drawing (Dissanayake, 2001; Matthews, 2003; Watanabe, 2012; D. Wolf & Perry, 1988). Temporal metrics and some spatial metrics based on fractal theory could indicate representativeness and anticipation, whilst the use of colours and space could indicate a sense of aestheticism. We used two datasets of drawings with different instructions in order to generalise results. We hypothesise that objective and quantitative patterns in drawings will provide cues about the intentions and representativeness of the drawer, even if the observer fails to perceive an object or entity in the drawing.

| Type | Metric | Definition | Meaning ('it measures') or expectation ('we expect') | References |
|---|---|---|---|---|
| Spatial metrics | µMLE | Spatial fractal metric, maximum estimate power coefficient of the drawing length distribution | It measures drawing efficiency, from random trajectories to optimal trajectories indicating a representativeness | (Edwards et al., 2007; Sueur, 2011; Sueur et al., 2011; G. Viswanathan et al., 2008) |
| | Drawing distance | Total distance of drawing, from the first point to the last, in pixels | We expect long distance drawings to be more representative or contain more details than short distance drawings. However, it can also mean deterministic drawing, (i.e no intention to represent anything). | (Handheld, 2020; Mitani & Nishida, 1993; Papi et al., 1995) |
| | Angle distribution metric | PCA dimension based on the coefficients of the cubic survival function of angle distribution (from 0° to 180°) | We expect homogeneous distributions of angles (low values of this metric) to indicate randomness (scribbles) whilst heterogeneous distributions should be linked to goal directedness (i.e. representativeness) | (Bartumeus et al., 2008; Benhamou, 2004; Gurarie et al., 2016; Potts et al., 2018) |
| | Minimum convex polygon | Minimum polygon covering all drawing points and giving the percentage of drawing cover on the screen | We expect high covering to inform about representativeness but also the play /emotional interest in drawing | (Gaston & Fuller, 2009; Nilsen et al., 2008) |
| Temporal metrics | Hurst index | Temporal fractal metric, measure of the long-term process in temporal sequence | It measures the temporal complexity of drawings sequences, from deterministic to complex | (A. MacIntosh, 2014; A. J. J. MacIntosh et al., 2011) |
| | Temporal Gini index | Measure of the inequality among values of temporal drawing sequences distributions | We expect high Gini Index, meaning unequal distribution of sequences to give an idea about intention and anticipation in drawing | (Debache et al., 2019; Planckaert et al., 2019) |
| | Entropy index | Measure of the temporal uncertainty of drawing | It measures the stochasticity or predictability of the drawing state. We expect a high entropy index to be linked to more representative drawing, including anticipation | (Ebeling et al., 2002; Kershenbaum, 2014; Leff, 2007) |
| | Drawing test time | Total drawing time (including drawing time and non-drawing time) | We expect a long duration to inform about thinking about drawing, i.e. intention and representativeness | |

|  | Metric | Definition | Meaning | References |
|---|---|---|---|---|
|  | Number of sequences | Number of drawing and non-drawing sequences during the test | We expect a high number of sequences to give an idea about goal-oriented behaviours, meaning intention and representativeness |  |
|  | Drawing speed | Speed of drawing, which is the drawing distance in pixels divided by the time of drawing | Speed is used as a measure of goal directedness or knowledge (i.e., in this context, mastering) | (Byrne et al., 2009; Noser & Byrne, 2014; Sueur, 2011) |
|  | Drawing time proportion | Drawing time divided by test time | We expect a high drawing time proportion to inform on thinking about drawing, i.e. intention and representativeness |  |
| Colour metrics | Mean colorimetric profile | Mean distribution of intensity levels for the Red, Green, or Blue colours respectively and after removing the white (screen) colour on the parts covered by drawing | It measures the mean spectrum of colours used, from dark to light. |  |
| Colour metrics | Standard deviation of the colorimetric profile | Standard deviation of the distribution of intensity levels for the Red, Green, or Blue colours respectively, on the parts covered by drawing | It measures the diversity of the spectrum of colours used. It is different from the number of colours as it takes also how much these colours are different. |  |
| Colour metrics | Number of colours | Number of colours used from the ten proposed colours. | We expect the number of used colours to give an idea about the aestheticism but also inform about the play interest in drawing. | (Martinet et al., 2021) |

Table 1: Metrics used to understand drawing complexity with their definition and their meaning in the context of drawing.

## 2. Material and Methods
### a. Dataset

*Dataset#1:* We asked 13 adults (6 men, 7 women), aged from 21 to 29 years old, to draw five drawings. These adults were considered "naïve" insofar that they had never taken drawing lessons and did not draw as a hobby. These participants were students of the research institute where the authors worked. Each drawing corresponded to an instruction that was specifically designed to produce a range of drawings: 1.) draw something with scribbles; 2.) draw something with circles; 3.) draw something with different angles; 4.) draw something with a starry sky; 5.) draw something with fan patterns (fan patterns are defined as straight lines with sharp angles due to repetitive actions of hand, usually from right to left). The reasons for the choice of these five instructions are detailed with the list of metrics in Section 2.c. Examples for each instruction are given in Figure 1. This dataset was collected in 2020 and is composed of 65 drawings.

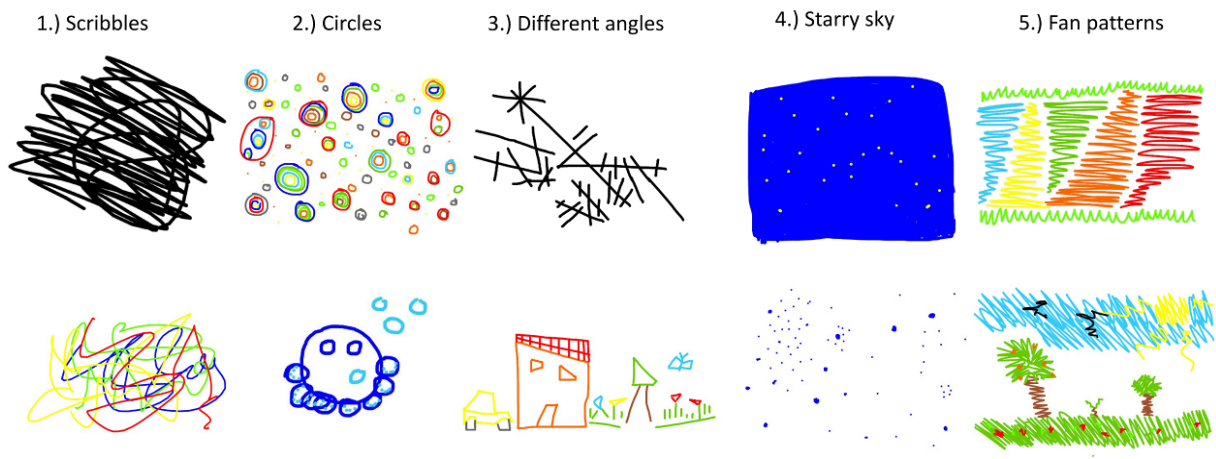

Figure 1: Examples of the five instructions we gave to participants for dataset#1. 1.) Make a drawing with scribbles; 2.) Make a drawing with circles; 3.) Make a drawing with different angles; 4.) Make a drawing with a starry sky; 5.) Make a drawing with fan patterns.

*Dataset#2:* This dataset included children and adults. The group of 144 children aged from three to ten years old was split into 18-20 children per one-year age interval. Boys and girls were equally represented in these categories except in the youngest category, which was composed of 5 girls and 15 boys. The adult group was composed of 41 adults (21 men, 20 women) aged 21 to 60 years old who were naive and expert drawers. The latter were art school students and professional illustrators. Participation was voluntary for adults and subject to parental consent for children.

All participants were asked to draw in two different conditions: free drawing (*draw what you want*: the experimenter told the subject that they could draw whatever they wanted, with no further instructions) and self-portrait conditions (*draw yourself*: the experimenter instructed the subject to draw themselves). The dataset was collected in 2018 and 2019. Further information about this dataset (i.e. methodology, examples of drawings and video footage of hand movements) is given in Martinet et al. (2021). A total of 370 drawings were initially collected for this dataset; however, some data were lost during the recording processes. The final dataset therefore contains 344 drawings. It should be noted that three of the naive subjects are present in both #dataset1 and #dataset2.

b. *Experimental design*

Habituation phase: each participant was invited to try a touchscreen tablet (iPad Pro, 13-Inch, version 11.2.2, capacitive screen reacting to the conductive touch of human fingers), and draw on it with their fingers to understand how it worked and how to change the colour they used to draw. Drawing with fingers was preferred in order to involve very young children who had not yet mastered the use of a pencil. A panel consisting of ten different colours was displayed at the bottom of the screen, and the participant could select a colour for their drawing by clicking on it. When they clicked on a different colour in the panel, any subsequent drawing production was in that colour. Children were habituated to the touchscreens the day before the tests to avoid overstimulation. Adults were tested immediately after discovering the tablet.

Testing phase: each child was individually tested during school time at school, located in their classroom (for the 3-year-olds) and in the staff room for the older children. The experimenter (LM or MP) stayed during the test but kept her distance during drawing to avoid influencing the children. Adults were tested individually in a room at the research institute (for naive participants) or at the art school (for expert drawers). Adult participants were left alone in the room. A camera recorded the hand movements of all participants while drawing, in case we needed to check for any problem during the session (interruption of the drawing, involuntary tracings, etc.). No time limit was applied.

c. *Data analysis*

For each drawing, the software developed for these studies (details and software available on demand), allowed us to record the spatial coordinates X and Y of every point of the lines drawn as well as their time coordinates [min; s; ms] and the colour used. This data collection allows us to calculate spatial, temporal and colour metrics per drawing (Table 1). Details of metrics, their calculation and the range of values for each instruction for dataset#1 can be found in the supplementary information section. The number of sequences is correlated to the number of lines in the drawing. We retain the number of sequences as data, as temporal sequences can be analysed using the Hurst index and analysed in parallel with the duration of each sequence.

For dataset#1, we expect different values according to the instruction:

1.) The scribbles drawings are not expected to show sharp angles and straight lines, so we predict the observation of a small µMLE (see Table 1) and a small angle distribution metric, but also a large minimum convex polygon. The drawing session duration and the number of sequences should be low but the drawing speed high. Finally, scribbles should have few colours.

2.) We expect that drawings with circles will not show sharp angles (but rather obtuse ones) or straight lines, so a small µMLE and a small angle distribution metric are predicted. We have no presupposition for the minimum convex polygon. Likewise, we cannot predict the drawing session duration or the number of sequences, whether in terms of speed or the number of colours used.

3.) We expect drawings with different angles to have large angle distributions as well as large distributions of lines lengths, meaning an intermediate µMLE. We have no hypothesis for the minimum convex polygon. The number of sequences should be high, corresponding to the different angles/lines but we cannot predict the duration of the drawing session, the drawing speed or the number of colours used.

4.) We expect drawings of a starry sky to contain different angles. We expect long distances between stars but short lines to draw stars, indicating a high µMLE. The minimum convex polygon, the number of sequences and the drawing session duration should all be high. The number of colours should be low and the colours should be light unless the participants drew a dark sky.

5.) We expect drawings with fan patterns to have sharp angles and long lines. We have no prediction for the minimum convex polygon. Speed should be high given the findings of literature on fan patterns (Kellogg, 1969; Zeller, 2007). The number of sequences should be high in relation to the different angles/lines. However, we have no prediction for the duration of the drawing session or for the number of colours used.

### d. Statistical analysis

As preliminary results, we analysed whether and how each metric differs between drawings for each instruction. This was achieved using ANOVA, or a Kruskal-Wallis test when ANOVA conditions could not be met (i.e., non-Gaussian distribution). Pairwise comparisons were realised when ANOVA or Kruskal-Wallis tests were significant (the "TukeyHSD" function of the R base package and the "kruskalmc" function of the "pgirmess" package (Giraudoux et al., 2018), respectively). Only differences with $p < 0.05$ were reported.

Analyses were carried out in three main steps using correlation analyses and principal component analyses: 1.) Analysis of dataset#1, 2.) analysis of dataset#2 following the same procedure as in step 1, 3.) a final combined analysis of dataset#1 and #2 in order to generalise our results.

First step on dataset#1: a correlation analysis was carried out with the R package "PerformanceAnalytics" (Carl et al., 2010; Peterson et al., 2018) on all metrics to identify those that were highly correlated. Following this correlation analysis, we removed the drawing duration proportion metric, which was highly correlated to the Gini index. Most of the variables were also influenced by the drawing test duration metric. We therefore decided to correct all the variables by carrying out a linear regression, using each metric as a response variable and the drawing test time as a factor. We took the residuals from this linear regression, which corresponds to any variance of each point that was not explained by the drawing test duration. A Principal Component Analysis (Budaev, 2010; Holland, 2008) with Varimax rotation was then carried out using the R package 'Psych' (Revelle, 2011; Revelle & Revelle, 2015). Variables are automatically corrected to be comparable (mean and range). Three dimensions were set up. Varimax rotation is used to simplify the expression of a particular subspace in terms of just a few major items each. This means that the Varimax rotation applies the variables to each dimension in turn in order to maximise the explained variance. We examined the loadings of each variable on each dimension. The loadings are interpreted as the coefficients of the linear combination of the initial variables from which the principal components are constructed. The loadings are equal to the coordinates of the variables divided by the square root of the eigenvalue associated with the component. We removed variables for which loadings are inferior to 0.4, which indicates a weak contribution to each dimension and to the total explained variance. After this removal, we renewed PCA with Varimax rotation and analysed the results.

Second step on dataset#2: We followed the procedure described for dataset#1.

Third step on dataset#1 and dataset#2: We compared the variables contributing to each of the three dimensions for dataset#1 and dataset#2. We removed the variables that did not contribute to the same dimensions between dataset#1 and dataset#2 and performed a PCA with Varimax rotation on both datasets. These results were then compared to assess whether our procedure might be generalised to any dataset. PCA dimensions were compared via a Pearson correlation test. Finally, the same PCA procedure was used to combine both datasets and compare the scribbles made by adults following the instruction we gave (« draw something with scribbles ») and the "natural" scribbles of 3-year-old children. A Mann-Whitney test was performed to compare both categories in each dimension.

All analyses were carried out using Rstudio 1.4.1103 (Allaire, 2012; Racine, 2012).

**Results**

Preliminary results on dataset#1: the details of tests and pairwise comparisons between instructions (dataset#1) for each metric are available in the supplementary material section. Three metrics showed similar values for the five instructions given to participants: angle distribution metrics, the mean colorimetric profile and the standard deviation of the colorimetric profile. The minimum convex polygon was lower in drawings composed of different angles than in fan pattern drawings. The number of colours was higher for the fan pattern instruction than in the different angles drawing. The scribble instruction results are different from all others in terms of drawing test time (i.e. lower), entropy (i.e. lower), number of sequences (i.e. lower), the Gini index (i.e. lower for all instructions except the starry sky) and the Hurst index (i.e. higher except in comparison to the fan pattern instruction). The starry sky instruction is linked to a longer drawing distance compared to all other instructions except fan patterns, whilst the different angles instruction is linked to a shorter drawing distance. The different angles instruction has a lower drawing time proportion and a higher Gini index than all other instructions except "draw circles".

First step on dataset#1: The results for the correlation analyses of metrics for the first dataset are shown in Figure 2. Drawing duration proportion is highly correlated (r=-1) with the Gini Index. We decided to remove drawing duration proportion as a variable. Moreover, as we could expect, ten of the 13 variables are correlated with drawing session duration. The latter is the most correlated with other variables. We then corrected all remaining variables according to the drawing duration. The correlation chart describing these corrected metrics is shown in Figure S15. This step was followed by a PCA with Varimax rotation. The total explained variance is 55.7% (Dimension 1 = 24.5%, Dimension 2 = 17.2%, Dimension 3 = 14%). Three variables have a loading below 0.4 for all three dimensions (details in Table S1, supplementary material), namely the angles distribution metric, the drawing session duration and the standard deviation of the colorimetric profile. We therefore removed these three variables from the dataset and carried out another Varimax rotation PCA. The total explained variance of this new PCA is 69.8% (dimension 1 = 31.9%, dimension 2 = 20.7, dimension 3 = 17.2%). Each metric shows a higher loading value in one dimension, unlike the two others (Table 2). We can thus attribute each metric to one dimension as follows: Dimension 1 (µMLE, drawing speed, Gini metric, entropy metric, drawing distance), Dimension 2 (minimum convex polygon, mean colorimetric profile, number of colours), Dimension 3 (Hurst index, number of sequences). Examples of dataset#1 drawings scaled to the three dimensions are given in Figure S16a-c.

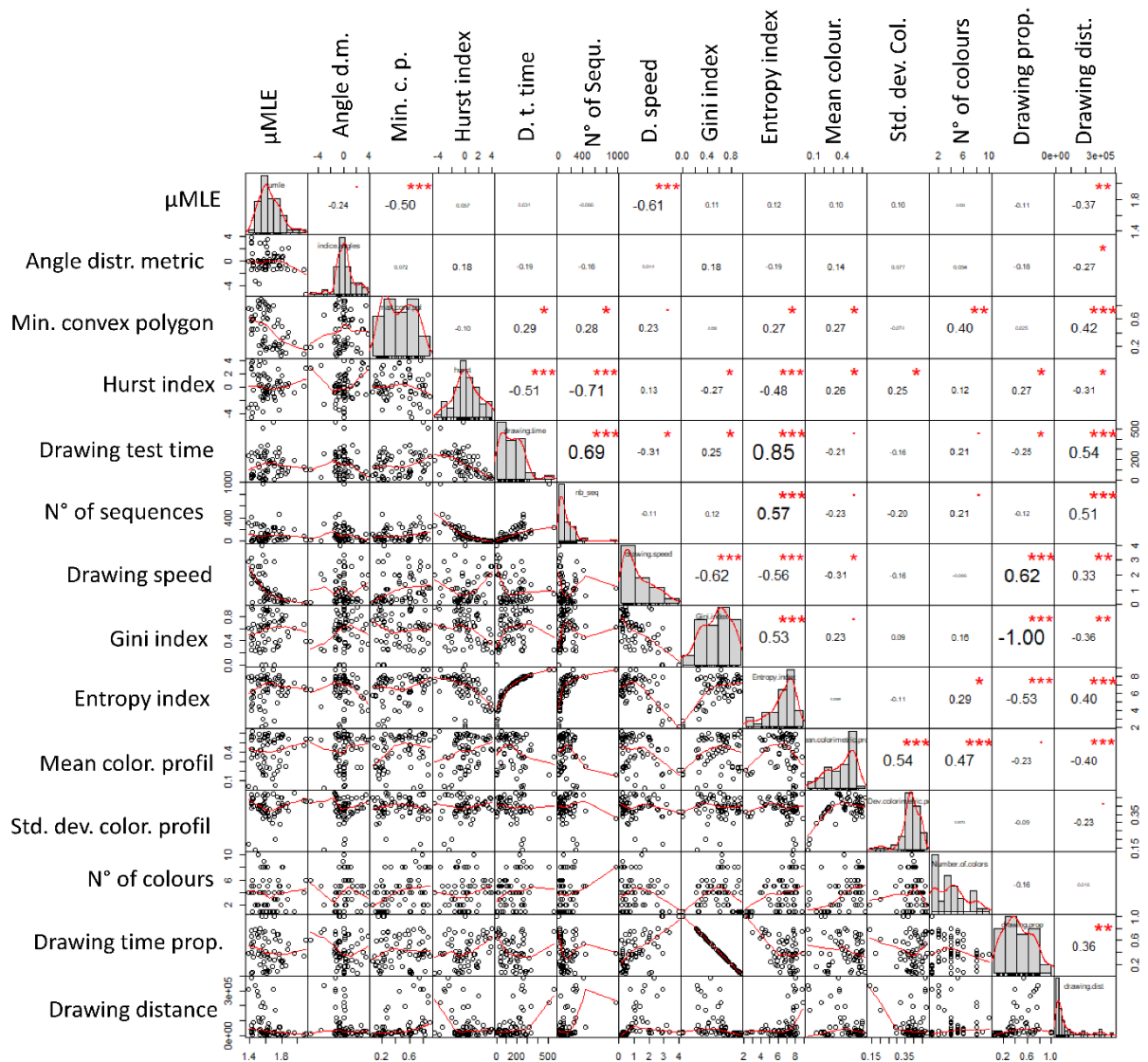

Figure 2: correlation chart of the 14 metrics for dataset#1. The diagonal of the graph provides the distribution of each metric, whilst the bottom left and the top-right provide the correlation figure and the correlation coefficient between two metrics, respectively. Statistical value is given with the correlation coefficient: * means p <0.05, ** means p <0.01, and *** means p <0.001.

|  | Dataset#1 | | | Dataset#2 | | |
|---|---|---|---|---|---|---|
|  | Dim. 1 | Dim. 2 | Dim. 3 | Dim. 1 | Dim. 2 | Dim. 3 |
| µMLE | **0.608** | -0.389 | -0.303 | **-0.781** |  | -0.276 |
| min. conv. pol | -0.374 | **0.783** | 0.134 | **0.666** | 0.404 | -0.133 |
| Hurst index |  | 0.214 | **-0.881** | -0.138 |  | **-0.911** |
| N° of sequences | -0.131 |  | **0.79** |  |  | **0.775** |
| Drawing speed | **-0.914** |  | 0.104 | **0.867** | -0.111 | 0.245 |
| Gini index | **0.784** | 0.19 | 0.211 | -0.14 | 0.42 | **0.685** |
| Entropy index | **0.659** | 0.363 | 0.212 |  |  | **0.45** |
| Mean colours profile | 0.437 | **0.711** | -0.191 | -0.168 | **0.857** |  |
| Number of colours | 0.11 | **0.759** | -0.144 | 0.181 | **0.607** |  |
| Drawing distance | **-0.754** |  | 0.223 | **0.708** | -0.488 | -0.296 |

Table 2: Loadings of the metrics (after loadings selection) on the three Varimax rotation PCA dimensions of dataset#1 and dataset2. Bold values indicate the dimension in which the metric is retained in each dataset. Grey highlights indicates similar results for both datasets. Axes of Dimension 1 are inversed between Dataset#1 and Dataset#2, but results and correlations are similar.

Second step on dataset#2: We followed the same steps as those described for dataset#1. Results for the correlation analyses of metrics of the second dataset are shown in Figure S17. The results of dataset#2 are comparable to those of dataset#1: the drawing duration proportion was highly correlated to the Gini index, and was therefore removed to correct other variables by the drawing duration. The correlation chart depicting these corrected metrics is shown in Figure S18. This step was followed by a Varimax rotation PCA. The total explained variance is 55.3% (dimension 1 = 18.7%, dimension 2 = 18.4%, dimension 3 = 18.2%). Like in dataset#1, the angle distribution and the drawing session duration have a loading below 0.4 for each dimension. However, the standard deviation of the colorimetric profile has a loading equal to 0.87 for dimension 1. We removed this variable to ensure a fit with the results of dataset#1; this does not change the variance explained (64.6% with versus 64.7% without) or the contributions of other metrics to the different dimensions. We carried out another Varimax rotation PCA. The total explained variance of this new PCA is 64.7% (dimension 1 = 24.2%, dimension 2 = 23.5%, dimension 3 = 17.1%). Each metric shows a loading value higher in one dimension, unlike the two others (Table 2). We can thus attribute each metric to one dimension as follows: Dimension 1 (µMLE, drawing speed, drawing distance, minimum convex polygon), Dimension 2 (mean colorimetric profile, number of colours), Dimension 3 (Hurst index, number of sequences, Gini metric, Entropy metric).

Third step on dataset#1 and dataset#2: Seven of the ten retained variables belong to the same dimension in the PCAs carried out for Dataset#1 and Dataset#2 (Table 2), and have quite similar loadings. However, three variables (minimum convex polygon, entropy index and Gini index) are not found in the same dimensions in the two datasets. When these three variables were removed from the PCA, we obtained similar results with comparable loadings per metric (Table 3) and 77.5% of the variance was explained for dataset#1 (dimension 1 = 31.9%, dimension 2 =23.2%, dimension 3 = 22.4%) , whilst 77% of the variance was explained for dataset#2 (dimension 1 = 31%, dimension 2 =

26.1%, dimension 3 = 19.9%). Reducing the selection of variables from ten to seven does not substantially change the classification of drawings, as the values of the three PCA dimensions are highly correlated between the first and the third step in dataset#1 (RC1: t = 18.942, df = 63, p <0.0001, r=0.92; RC2: t = 14.357, df = 63, p <0.0001, r=0.87; RC3: t = -15.764, df = 63, p <0.0001, r=0.89). When we combined both datasets, we obtained similar results to those obtained in separate analyses of dataset#1 and dataset#2, with 77.5% of the variance explained (dimension 1 = 30.2%, dimension 2 =25%, dimension 3 = 20.2%; Table 3). Examples of dataset#1 drawings scaled on the three dimensions are given in Figure 3a-c. Finally, we compared the three dimensions between scribbles of dataset#1 (made by adults) and scribbles of dataset#2 (made by 3-year-old children only, as scribbles become rare from the age of four onwards). Mann-Whitney tests showed that Dimension 1 differs between dataset#1 and dataset#2 (w=29, p=0.0066) whilst there is no significant difference between the two sets for dimension 2 (w=107, p=0.122) and dimension 3 (w=105, p=0.152) (Figure 4). Moreover, Figure 4 shows that data are more dispersed for the three dimensions in adults' scribbles compared to children ones. Mann-Whitney test for each metrics in each dimension are detailed in the supplementary material (Table S2).

|  | Dataset#1 | | | Dataset#2 | | | Dataset#1 & #2 | | |
|---|---|---|---|---|---|---|---|---|---|
|  | Dim. 1 | Dim. 2 | Dim. 3 | Dim. 1 | Dim. 2 | Dim. 3 | Dim. 1 | Dim. 2 | Dim. 3 |
| µMLE | **0.812** |  |  | **0.814** | -0.117 | 0.246 | **0.778** | -0.14 | 0.252 |
| Hurst index |  | 0.248 | **0.872** | -0.143 |  | **0.902** |  |  | **0.905** |
| N° of sequences | -0.188 | 0.112 | **-0.886** |  |  | **-0.898** |  |  | **-0.896** |
| Drawing speed | **-0.894** | -0.133 |  | **-0.9** |  | -0.216 | **-0.88** |  | -0.136 |
| Mean colours profile | 0.317 | **0.81** |  | 0.291 | **0.777** |  | 0.315 | **0.777** |  |
| N° of colours |  | **0.909** |  | -0.178 | **0.795** |  | -0.141 | **0.824** |  |
| Drawing distance | **-0.793** | -0.194 |  | **-0.746** | -0.371 | 0.296 | **-0.775** | -0.312 | 0.21 |

Table 3: PCA loadings of the metrics similar to dataset#1 and dataset#2. Bold values indicate dimensions in which the metric is retained in each dataset.

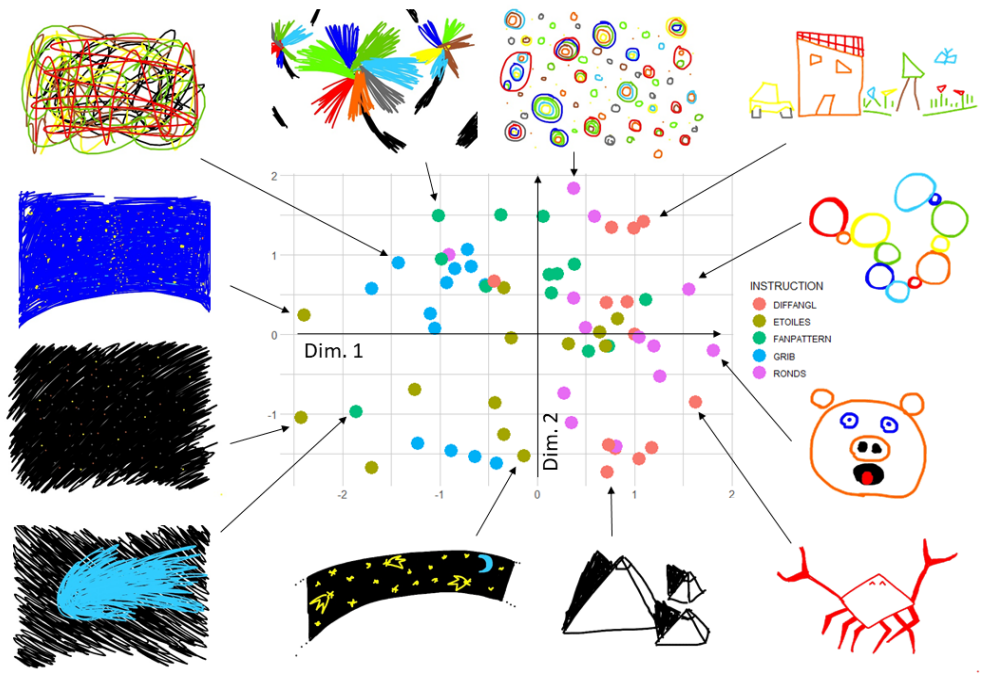

Figure 3a: Examples of Dataset#1 (third step) drawings according to Dimension 1 and Dimension 2, as provided by the PCA. Dimension 1 may represent representativeness in drawing whilst Dimension 2 may represent diversity in drawings.

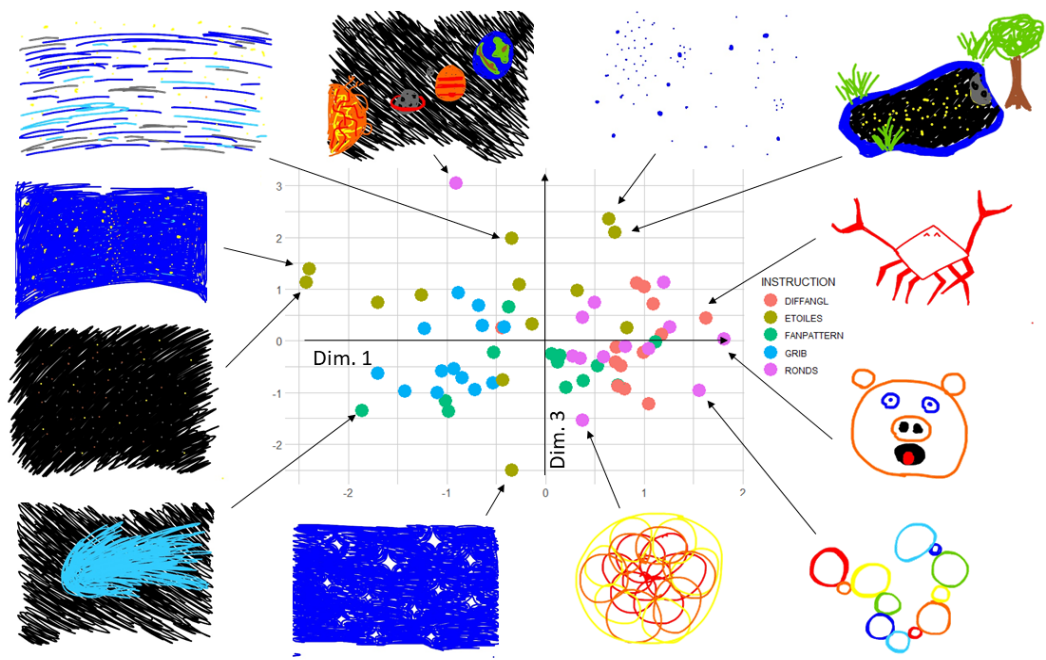

Figure 3b: Examples of Dataset#1 (third step) drawings according to Dimension 1 and Dimension 3, as provided by the PCA. Dimension 1 may represent representativeness in drawing whilst Dimension 3 may represent periodicity in drawings.

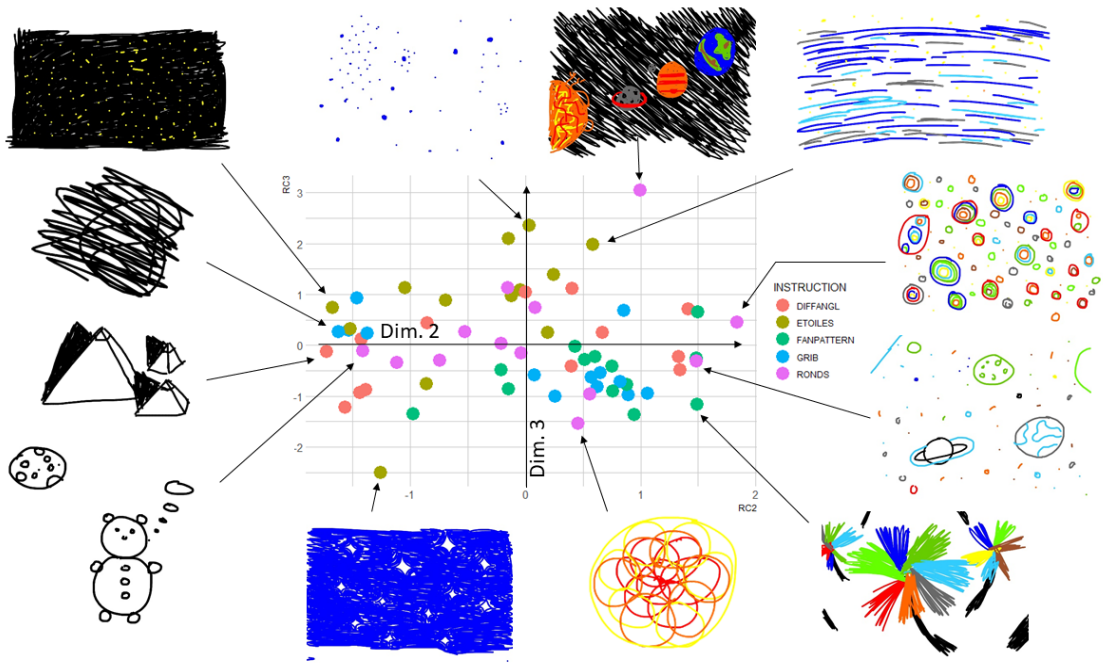

Figure 3c: Examples of Dataset#1 (third step) drawings according to Dimension 2 and Dimension 3, as provided by the PCA. Dimension 2 may represent diversity in drawing whilst Dimension 3 may represent periodicity in drawings.

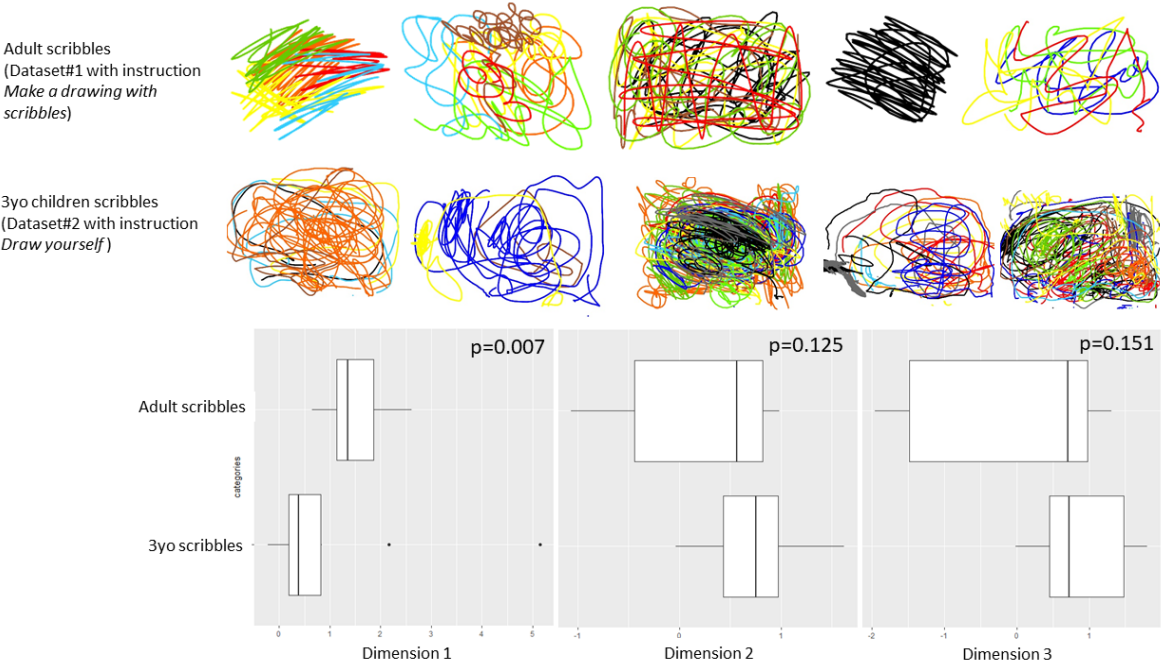

Figure 4: Examples of scribbles for dataset#1 (adults) and dataset#2 (3-year-old children) and boxplot of dimensions 1 to 3 for these two categories.

**Discussion**

We used several mathematical metrics to characterise drawings and assess whether they can give cues about representativeness and intention. Principal component analyses helped us to organise these metrics in three dimensions in a first dataset. The analyses on this first dataset were then confirmed with the analyses carried out on a second dataset, thus allowing us to generalise our method of characterising drawings and the subsequent results. This study is an important step in the analysis of drawings as it is the first time that such a high number of mathematical indices are used to analyse the cognitive processes behind creativity. This discussion seeks to understand which process corresponds to each dimension provided by the PCA.

The choice of two datasets that each involve different drawing instructions proves to be the right protocol to obtain variations in each metric. The colours used to draw are influenced by the instructions we gave. The choice of these instructions was designed to produce a variety of shapes and lines influencing spatiotemporal metrics. However, variation in used colours is still high and two colour metrics (i.e. number of colours and mean colorimetric profile) partly explains variance in different dimensions. Nevertheless, the standard deviation of the colorimetric profile does not provide any information about the drawing, as the choice of colours variable is more a question of personal preference than characteristic of any representative process. Similarly, and contrary to what we expected, the angles distribution metric does not differ between the instructions of dataset#1 and did not play any role in explaining variance in the PCAs of dataset#1 and dataset#2. This result might be due to the issuing of an unsuitable instruction, thus leading to false negatives. However, as we obtained similar results between dataset#1 and dataset#2, the explanation should be more in the drawing process itself where, whatever the objects, their representation produces similar angle distributions. The study of turning angles is often used when assessing the optimality of animal movements, but it is limited to differentiating random movements from goal-oriented and directed ones (Reynolds, 2008; Sueur, 2011; Sueur et al., 2011). The scale used to analyse angles in our study is possibly too limited to obtain significant results. After correcting all variables according to the drawing session duration, no difference is seen in the latter between instructions, nor did it play a role in explaining PCA variance in either of the datasets. This means that representativeness and aestheticism in drawing are not directly linked to drawing duration but more to all the other processes (e.g. the drawing traits length and the number of sequences depending on the number of objects) that influence the duration of drawing. In Dataset#1, scribbles were the only instruction leading solely to non-representative drawings, despite the fact that the definition of a scribble can be unclear. Other instructions mainly resulted in the drawing of objects or animals. It is difficult to assess whether it is the instruction itself or the process of drawing scribbles that leads to non-representative drawings, but scribbles show the highest difference with other instructions for many metrics in dataset#1. Although it is possible to draw a figurative drawing using scribbles, no such cases were observed in our study. This is certainly because participants had toddler scribbles in mind when we gave the instruction. Observation of these differences shows that the process of drawing scribbles results in the rapid drawing of a small number of short sequences of relatively random lines.

Closer evaluation of the principal components analyses shows that some metrics obtained a high loading but were not found in the same dimensions in dataset#1 and dataset#2. This was the case for the minimum convex polygon, which does not show substantial differences between the instructions of dataset#1. We would expect the minimum convex polygon to be a proxy of representativeness or aestheticism by filling the screen, but toddlers are reported to fill the paper sheet when drawing (Kellogg, 1969; Matthews, 1984; D. Wolf, 1988). Indeed, the minimum convex polygon was also shown to be high with scribbles or other non-representative drawings (see

Figure 3), which indicates that this metric cannot be used to understand the cognitive processes underlying drawing. The Gini index and entropy index, both measured on temporal sequences of drawing, also belong to different dimensions in dataset#1 and dataset#2. The Gini index is a measure of the inequality of temporal drawing sequences, whilst entropy is a measure of the temporal uncertainty of drawing. However, the duration of drawing sequences is linked to the lengths of the drawing lines for each object in the drawing. This can be seen in the different correlation charts, where temporal metrics are correlated to spatial ones. Given this uncertainty in the explanation of the dimensions of dataset#1 and dataset#2, we preferred to remove these two metrics. However, PCA results did not change after this removal and the explained variance was higher. The Gini index and entropy index are increasingly used in different studies, despite a continuing debate about their interpretability (Ben-Naim, 2012; Leff, 2007; Lerman & Yitzhaki, 1984). Future works are required to assess their potential role in the domain of drawing and other behaviours. Moreover, the Gini index is ineffective when calculated on binary sequences and does not provide any new information. Perhaps considering the cumulative sum of drawing and non-drawing could lead to meaningful results.

After the different steps of selection of variables, we reached similar results between dataset#1 and dataset#2 and explained almost 80% of variance. This result is important because it means that whatever the dataset and the given instructions, our method could be applied to analyse drawings. However, this method is only valuable if the dimensions of the PCA have a biological or psychological aspect.

Dimension 1 is composed of the drawing speed, the drawing distance and the µMLE (see Table 1 for definitions). In movement ecology, the drawing speed is a proxy of goal directedness, i.e. the intention of an animal to go to a specific place that it knows (Byrne et al., 2009; King & Sueur, 2011; Noser & Byrne, 2014; Sueur, 2011). The higher the motivation to go to a place where resources can be found, the higher the speed. In our drawing context, speed can be a proxy of intentionality and of mastering, meaning that participants who are familiar with what they are drawing do it faster. We observed this tendency in dataset#2, where the mean drawing speed of experts was 0.73±0.37 compared to 0.56±0.27 for naive participants. However, drawing speed is also high for scribbling toddlers or for someone who wants to fill the screen (or paper sheet) with one colour, and in both of these cases, drawing speed is linked to drawing distance. This is what we obtained in our results (Tables 2 & 3, Figure 3) with many participants colouring the starry sky blue or black. This adds detail without providing a better representativeness of the drawing. On the other hand, µMLE is negatively correlated to drawing speed and drawing distance, and is used in ecology to evaluate the efficiency of animal trajectories. In an environment with different food resources, an animal can move randomly or go directly to the resources area. In this case, movements are efficient and optimal and µMLE is high. We qualified these movements as a Lévy flight (or walk) (Reynolds, 2008; G. M. Viswanathan et al., 1999). In our drawing study, a high µMLE indicates an efficient drawing, i.e. it is representative, intentional and with few details (see Figure 3). It is easy to recognise what the drawer wanted to draw, but drawing distance is low and indicates few details. If we could link an ability to this Dimension 1, it would be efficiency. In this way, Dimension 1 can be named efficiency. Indeed, efficiency can be defined as an ability to avoid wasting materials, energy, efforts, money, time, etc. Efficiency is different from effectiveness, which is the capability of producing a desired result or the ability to produce desired output (Frøkjær et al., 2000; Marley, 2000). In our case effectiveness is representativeness, whilst efficiency is representativeness combined with few details (i.e. optimality). Efficiency might be illustrated in a sketch (Mihai & Hare, 2021; Xu et al., 2020) and by emoticons (Huang et al., 2008; Takahashi et al., 2017). Importantly, the scribbles made by adults showed higher values of Dimension 1 than scribbles by toddlers. This is particularly due to higher drawing speed in adults (Table S2). The number of sequences is also higher for adults' scribbles, but the number of colours is lower. This may indicate that even if these drawings do not have an external representativeness, they may have an internal representativeness for adult participants.

The second PCA dimension is composed of the number of used colours and the mean colorimetric profile. Adding and diversifying colours facilitates the differentiation of objects in a drawing. When there are few details in a drawing, one colour is enough to identity the object but when more details are present, the use of colours makes it easier to identify the different objects. This principle can be observed in Figure 3a, for instance, with the drawings of the crab (one colour) and the house (different colours to identify the flowers, the car, the butterfly, etc.). Colours facilitate the visual perception of objects and materials in our environment (Castelhano & Henderson, 2008; Witzel & Gegenfurtner, 2018). In our drawing datasets, this cognitive process is found to make drawings easier to interpret and increase their external representativeness. This second dimension can be named "diversity" to represent the diversity of colours in terms of number and panel.

Finally, the third PCA dimension is composed of the Hurst index and the number of sequences, both of which are temporal metrics. The number of sequences is directly dependent on the number of lines drawn (whatever their length), i.e. the number of forms that are either objects or components of an object. The Hurst index is a proxy of the temporal complexity of a drawing. It indicates how far the timeline of a sequence can predict another sequence. For instance, two drawings can contain the same number of sequences, but one will be considered as deterministic (small values of dimension 3, not complex, for analyses of Dataset#1) if the duration of sequences is similar (because the drawn objects are all similar), whilst the other will be considered stochastic (high values of Dimension 3) and more complex if the sequences cannot be predicted because they followed an unpredictable pattern (which is the intention when representing different objects). Such examples can be seen for instance in Figures 3b and 3c (a drawing that resembles a rose window and the drawing with planets) for Dataset#1. Here, higher values in Dimension 3 seem to indicate higher anticipation and intention in drawing. Dimension 3 can be named "sequentiality". When the complexity of the drawing increases to make something representative, the number of sequences and the stochasticity increase.

Our study showed that we can identify three dimensions in drawing: efficiency, diversity and sequentiality. All three dimensions facilitate our understanding of drawing representativeness. This statement can be observed in Figure 3, where we can easily determine the intentions of participants, even if the drawing is abstract (i.e. there are no identifiable objects). The combination of these three dimensions allows us to judge the representativeness of a drawing even if it does not seem to represent anything for the observer. The perspectives of this study are therefore noteworthy: we can use this method to evaluate the intentions behind a drawing that has no meaning for us, as adults with no psychopathologies. In this respect, we identify three perspectives: 1) We can study the ontogeny of drawing in children and identify with precision the premises of the different steps observed during the drawing learning process (action representation, romancing and guided elicitation). 2) We can also extend the study to other species, particularly great apes, which are known to draw, and assess whether their drawing is motivated by internal representativeness (Martinet & Pelé, 2020). 3) Finally, the method can be extended to psychopathologies such as autism (Charman & Baron-Cohen, 1993; Jolley et al., 2013) and even certain emotional disorders (Desmet et al., 2021; Nolazco-Flores et al., 2021), or simply be used to measure learning difficulties and creativity (Lee & Hobson, 2006; Urban, 2004). New technologies combined with new mathematical methods appear to be very useful and provide new possibilities to test mental states, intentions and emotions beyond these representations (Watanabe & Kuczaj, 2012). However, the meaning of each dimension has been assessed in visual terms only and is therefore not completely objective. New metrics could ultimately lead to the discover new cognitive dimensions and meanings, or reinforcing the discoveries of this study. Finally, the participants who have drawn for this study are all from France. To make these results universal, it could be useful to collect drawings from around the world.

**Acknowledgments**

We thank the director and the teachers of the school for giving us access to their classrooms and showing interest in our research project. We are grateful to all the participants and to the parents of all the children, who accepted with enthusiasm to contribute to our study. Thanks also to Sarah Piquette, who provided help for the ethical components of this project.

**Declarations**

Funding: This project has received financial support from the CNRS through the MITI interdisciplinary programs and an IDEX Exploratory Research program from Strasbourg University.

Conflicts of interest: The authors declare having no conflicts of interest.

Ethics approval: Study protocol followed the ethical guidelines of our research institutions and ethical approval was obtained from the Strasbourg University.

Consent to participate: Informed consent was obtained from all adult participants and from a parent or legal guardian for children. Informed consent for publication of identifying images in an online open-access publication has been obtained too.

Research Ethics Committee (Unistra/CER/2019-11).

Data availability: The datasets generated during and/or analysed during the current study are available in the Zenodo repository, https://doi.org/10.5281/zenodo.5387520

# Supplementary material for

# Making drawings speak through mathematical metrics.

Cédric Sueur[1,2], Lison Martinet[1], Benjamin Beltzung[1], Marie Pelé[3]

1: Université de Strasbourg, CNRS, IPHC UMR 7178, Strasbourg, France

2: Institut Universitaire de France, Paris, France

3 : Anthropo-Lab, ETHICS EA7446, Lille Catholic University, Lille, France

Corresponding author: Cédric Sueur, cedric.sueur@iphc.cnrs.fr; IPHC UMR 7178, 23 rue Becquerel 67087 Strasbourg, France

**Details about metrics, their calculation and a value range for each instruction in dataset#1**

**1. µMLE**

µMLE is the maximum likelihood exponent that results from the analyses of the drawing-line length distribution. This analysis originates from the random walk or Lévy walk theory (Sueur, 2011; Sueur et al., 2011). The random walk analysis determines whether the distribution of drawing lines follows a power law or an exponential law. If the distribution follows an exponential law, we expect the drawing to be random, meaning that the individual who is drawing does not intend to represent any specific thing. In contrast, a power distribution should reflect a non-random and oriented behaviour, as found for the daily paths of animals in their natural environments (i.e. goal-oriented and efficient movements; Sueur, 2011; Sueur et al., 2011). Details about the use of this metric for drawing analysis are developed and detailed in (Martinet et al., 2021).

As the coordinate scoring of the drawing was continuous (one point per frame), we focused on active changes (Byrne et al., 2009; Noser and Byrne, 2014): a selection of points was carried out for each drawing via a change-point test under R software (version 1.1.383; CPT, script available in Noser and Byrne, 2014). Two consecutive points (i and j) in the drawing determined a step or a vector of a length L (i, j). We then calculated the step lengths S on Excel with latitude x and longitude y (in pixels). Step lengths between 0 and 10 pixels were removed since they often corresponded to very short, inactive movements such as imprecise lines or finger sideslips, and caused inaccuracies. We then determined whether the step length frequency distribution of a drawing followed a power law ($y = a*x^{\mu}$) or an exponential law ($y = a.e^{x*\lambda}$) using the Maximum Likelihood Method (Edwards et al., 2007; Martinet et al., 2021). Log-likelihood of the exponential and power distributions for each drawing could then be compared using the Akaike Information Criterion (AIC). We retained the model retained (power or exponential) with the lowest AIC, with a minimum difference of 2 between the two AICs (Burnham and Anderson, 2004). All the drawings produced followed a power law. The Maximum Likelihood Estimate of the power law exponent µMLE was then used to draw conclusions on the efficiency of the representation for each drawing. This index is comprised of values between 1 and 3. The higher the index, the more the line was considered to be directed, well planned and efficient (Bartumeus et al., 2005; Viswanathan et al., 1996).

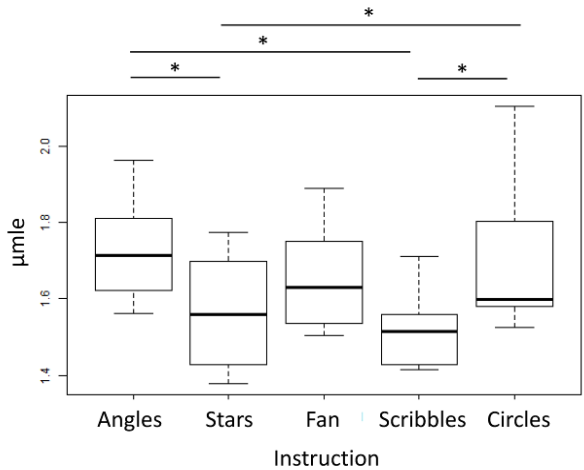

Figure S1: Boxplot of the µMLE according to instructions. Stars indicate significant difference between two instructions with p <0.05. ANOVA test gives a p= 0.0003.

**2. Drawing distance**

Drawing distance is the total distance of drawing in pixels, from the first point to the last. Two consecutive points (i and j) in the drawing determined a step or a vector of a length L (i, j). We then calculated the step lengths S on Excel with latitude x and longitude y (in pixels). We calculate the sum of all lengths S as the drawing distance.

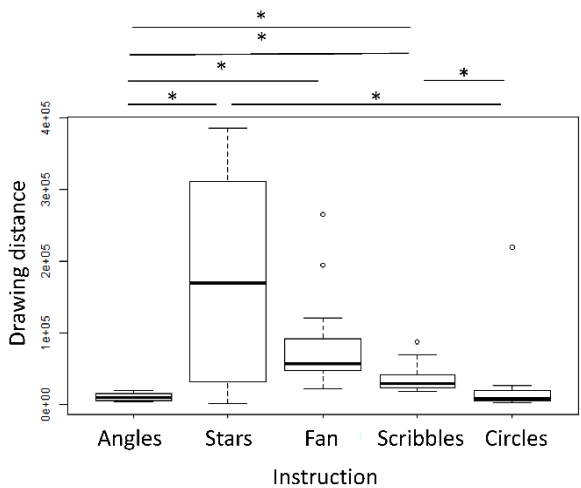

Figure S2: Boxplot of the drawing distance according to instructions. Stars indicate significant difference between two instructions with p <0.05. Kruskal-Wallis test gives a p-value <0.0001.

**3. Angle distribution metric**

Goal directedness in animals is assessed by individuals moving in straight lines with limited tortuosity, except when they arrive at a food resource site and start to forage. Drawing intentionality can be considered similar to animal food research efficiency (Martinet et al., 2021; Martinet and Pelé, 2020). Analyses of angle distributions (i.e. turning angles) between two trajectories or lengths can be considered as a reliable way to measure goal directedness and tortuosity (Bartumeus et al., 2008; Hurford, 2009; Potts et al., 2018). A turning angle is the difference in direction for two successive vectors or steps. We followed the same methodology as described in **1. µMLE.** As coordinate scoring of the drawing was continuous (one point per frame), we focused on active changes (Byrne et al., 2009; Noser and Byrne, 2014): a selection of points was carried out for each drawing via a change-point test under R software (version 1.1.383; CPT). Step lengths between 0 and 10 pixels were removed as they often corresponded to very short, inactive movements such as imprecise lines or finger sideslips, and caused inaccuracies. Function ATAN2 in Excel was then used to calculate the angle (in radians) between two consecutive points and convert radians into degrees. We applied corrections to only retain angles between 0° and 180°, as movements are oriented. We then calculated the survival distribution of angles (i.e. going from 1 or 100% of points to 0) for each drawing. A cubic function was fitted to the distribution. Examples of drawings and the respective distributions are given in Figure S3.

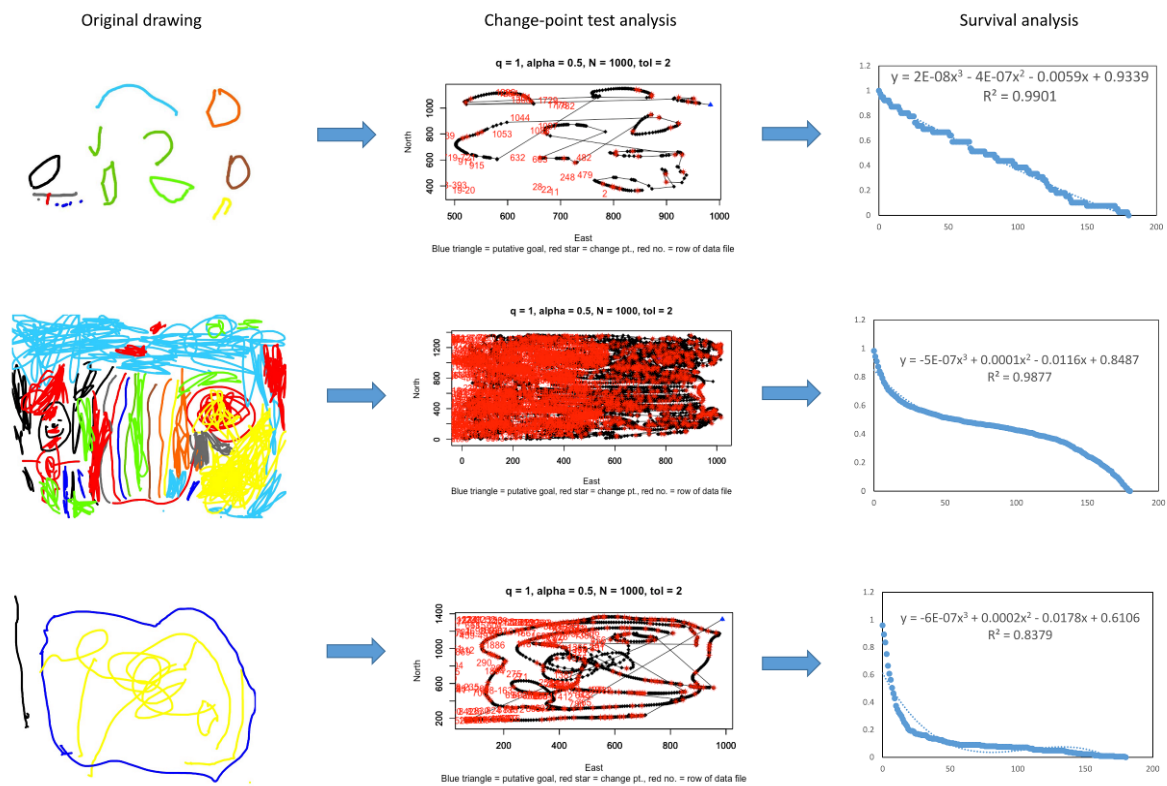

Figure S3: Analysis of angles distribution. The original drawing is transformed into vectors (i.e. trajectories) following the change-point test, and angles are calculated. The angle survival distribution is then fitted with a cubic function.

The cubic function is y=-ax$^3$+bx$^2$-cx+d. The more the distribution looks like a straight line, the lower the values of a, b and c will be. The more the curve looks like a sigmoid or inversed power function, the higher the values of a, b and c will be. The three constants a, b and c are highly correlated and can be combined using a principal component analysis, where one dimension explained 82% of variance for dataset#1 (dimension 2 = 15%) and 90% of variance for dataset#2 (dimension 2 = 5%). We used the values of dimension 1 as values for the angle distribution metric for each drawing.

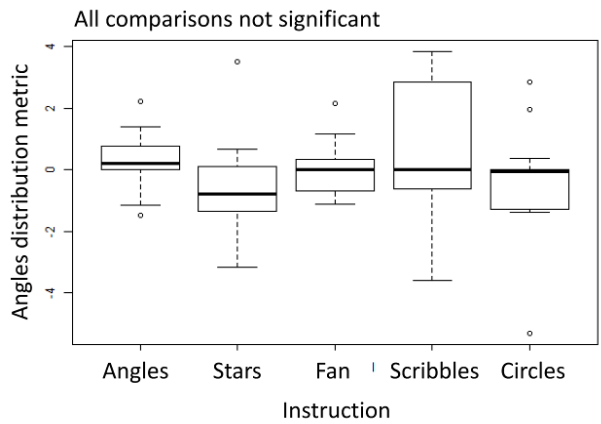

Figure S4: Boxplot of the angle distribution metric according to each instruction. ANOVA testing provides a P value of 0.329

### 4. Minimum convex polygon

The minimum convex polygon draws the smallest polygon around points with all interior angles measuring less than 180 degrees. Minimum convex polygons are common estimators of home range (Nilsen et al., 2008) but represent the cover of drawing on the screen, between 0 (no drawing at all) to 100% (the drawing covered the entire screen). We used the scissors select tool in GIMP 2.10.22 (Lecarme and Delvare, 2013; Peck, 2006) to select the minimum convex polygon of drawings.

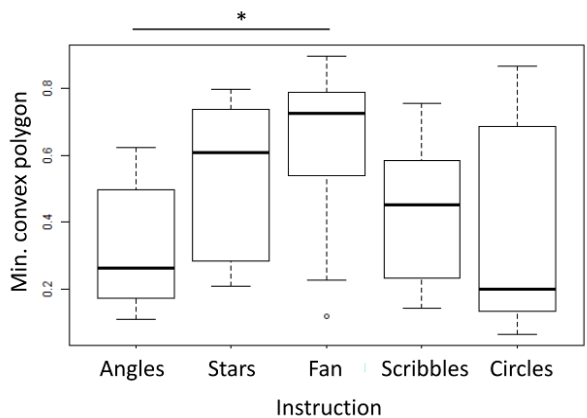

Figure S5: Boxplot of the minimum convex polygon according to instructions. Stars indicate significant difference between two instructions with p <0.05. ANOVA testing provides a P value of 0.016.

### 5. Hurst index

Details about the calculation of this metric can be found in Beltzung et al. (2021). Fractality or long-term processes can be measured by different methodological approaches, each of which has its own fractal statistical parameter. Here, the difficulty lies in the fact that numerous estimators have been defined for each parameter, yet the effectiveness of these very estimators is still debated in the literature (Stadnitski, 2012; Stadnytska et al., 2010). Studies often focus on one or a small number of estimators without a rigorous reason (such as comparing them). As a consequence, there is no simple and systematic way to estimate the long-memory process, which often results in errors or misleading conclusions in studies (Karagiannis et al., 2006). The most widespread way to assess and quantify long-memory processes in temporal sequences is the estimation of the Hurst exponent H. Indeed, a behavioural state is influenced by previous states following two different scenarios: *persistence*

(H>0.5) when a positive correlation occurs, meaning that a long sequence is likely to be followed by a long sequence in the future, and *anti-persistence* (H <0.5) when a long sequence is likely to be followed by a short sequence, i.e. a negative correlation (Delignières et al., 2005). Here, we combined different methods using a PCA as explained in Beltzung et al. (2021).

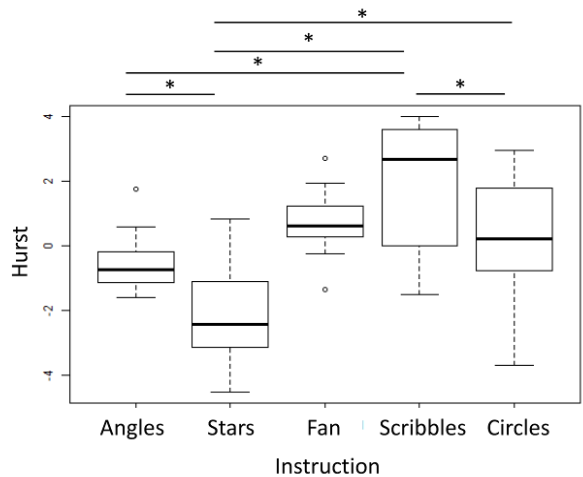

Figure S6: Boxplot of the Hurst index according to each instruction. Stars indicate significant difference between two instructions with p <0.05. ANOVA test gives a p <0.0001.

**6. Temporal Gini index**

The Gini index was calculated on the binary sequences (drawing / non-drawing) and describes the uniformity of the distribution. This index is a real number between 0 and 1, where a value of 0 indicates perfect equality, and a value of 1 indicates maximal inequality. The interpretation of this index on behavioural sequences can be questioned, as well as the temporal sequence to consider. The Gini index was calculated by using the R package "DescTools" (Signorell et al., 2016).

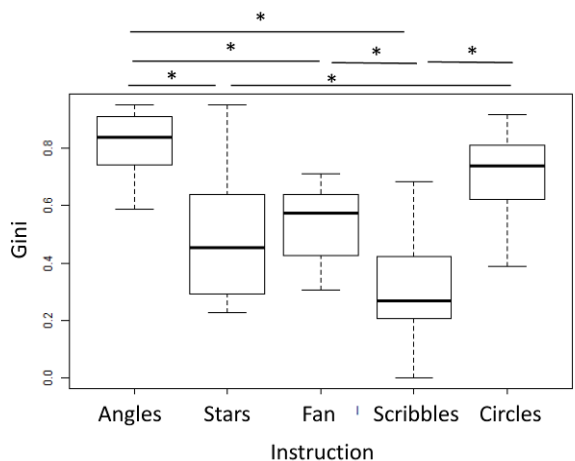

Figure S7: Boxplot of the Gini metric according to instructions. Stars indicate significant difference between two instructions with p <0.05. ANOVA test: p <0.0001.

**7. Entropy index**

The Shannon entropy index (Gray, 2011; Leff, 2007) has not been calculated for temporal binary sequences, but for the cumulative sum of drawing time at each second. This index characterises the quantity of information contained in a variable. The higher the quantity of information, the higher the uncertainty (the entropy). For a string of sequences with n distinct sequences, each sequence has a frequency of p. The entropy index was calculated by using the R package "DescTools" (Signorell et al., 2016). The entropy of **Shannon** H is calculated according to the formula:

$$H = -\sum_{i=\&}^{n} p_i \log_2 p_i$$

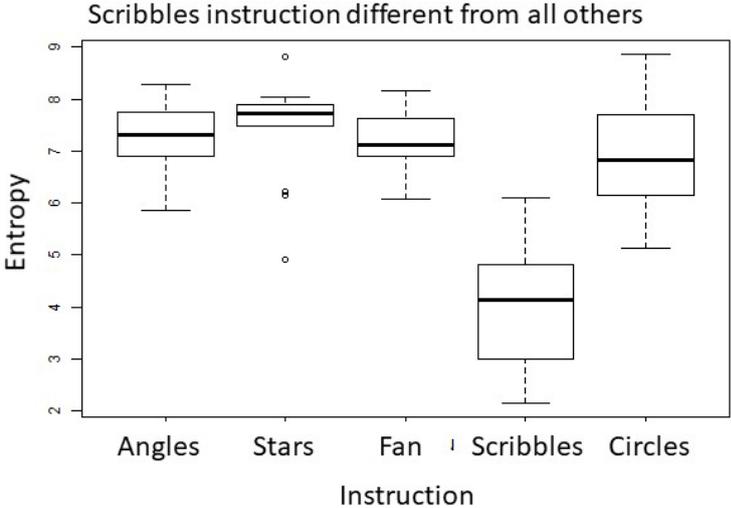

Figure S8: Boxplot of the entropy metric according to instructions. Stars indicate significant difference between two instructions with p <0.05. ANOVA test: p <0.0001.

**8. Drawing duration**

The duration of the drawing session is defined as the time from the first point on the touchscreen to the last, including the time spent drawing and time without drawing between these two points.

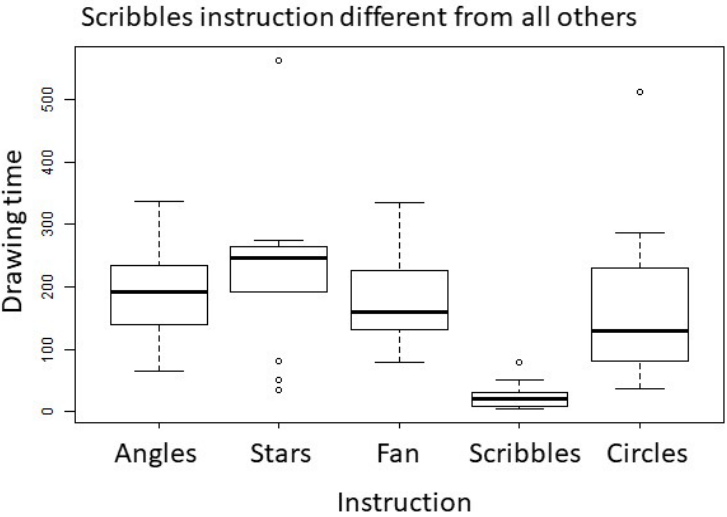

Figure S9: Boxplot of the drawing duration according to instructions. Stars indicate significant difference between two instructions with p <0.05. ANOVA test: p <0.0001.

## 9. Number of sequences

Number of drawing and non-drawing sequences during the test.

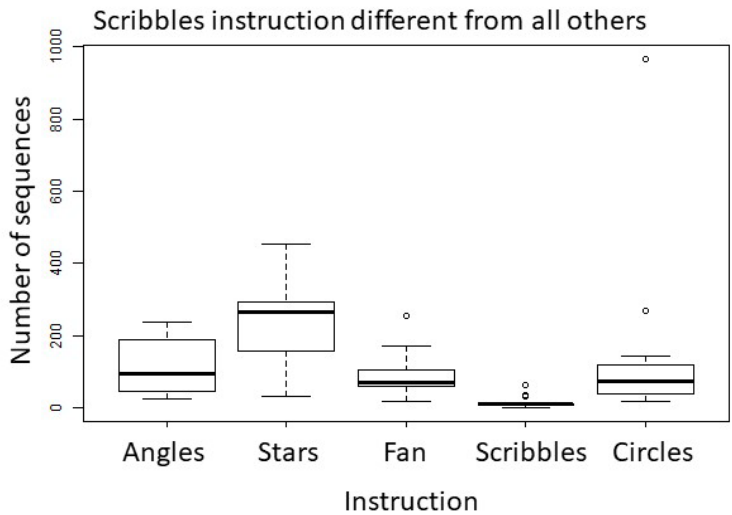

Figure S10: Boxplot of the number of sequences according to instructions. Stars indicate significant difference between two instructions with p <0.05. Kruskal-Wallis test: p <0.0001.

## 10. Drawing speed

Drawing distance divided by the drawing duration (i.e. duration of drawing only, not including duration of non-drawing during the test).

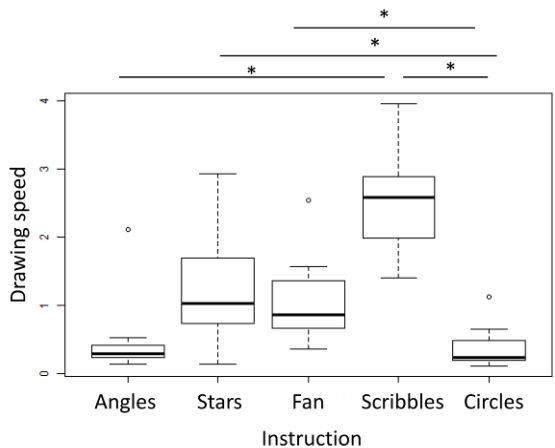

Figure S11: Boxplot of the drawing speed according to instructions. Stars indicate significant difference between two instructions with p <0.05. ANOVA test: p <0.0001.

## 11. Drawing time proportion

Duration of drawing divided by the duration of drawing session.

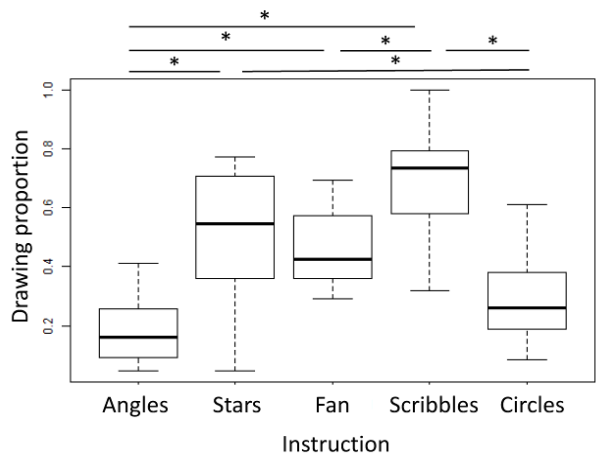

Figure S12: Boxplot of the drawing proportion according to instructions. Stars indicate significant difference between two instructions with p <0.05. ANOVA test: p <0.0001.

## 12. Mean colorimetric profile

Mean distribution of intensity levels for the Red, Green, or Blue colours respectively and after removal of the white (screen) colour on the parts covered by drawing (after selection with the minimum convex polygon).

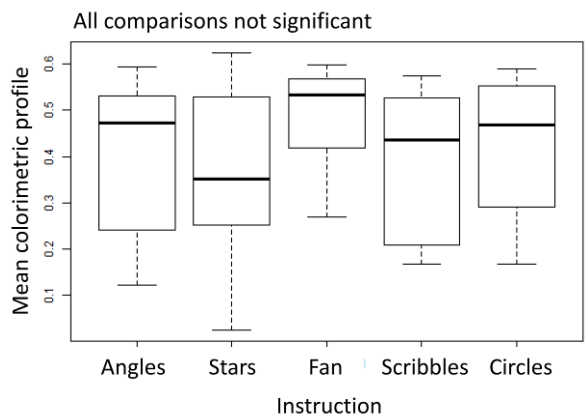

Figure S13: Boxplot of the mean colorimetric profile according to instructions. Stars indicate significant difference between two instructions with p <0.05. ANOVA test: p=0.319.

## 13. Standard deviation of the colorimetric profile

Standard deviation of the distribution of intensity levels for the Red, Green, or Blue colours respectively, on the parts covered by drawing (after selection with the minimum convex polygon).

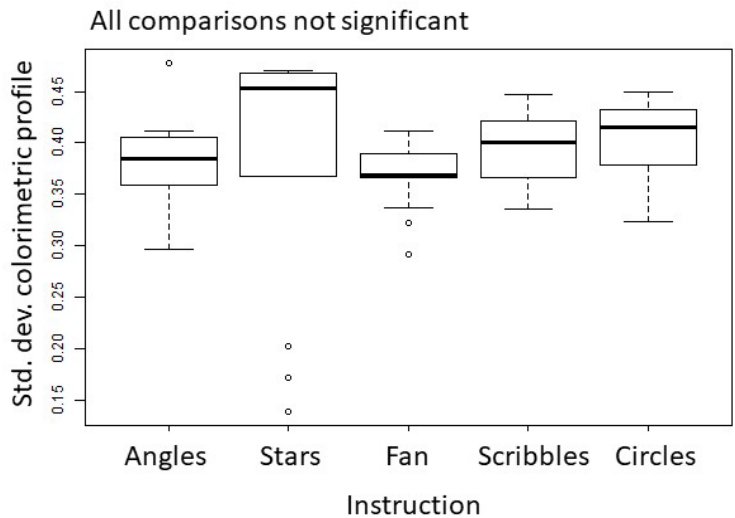

Figure S14: Boxplot of the standard deviation of the colorimetric profile according to instructions. Stars indicate significant difference between two instructions with p <0.05. ANOVA test: p=0.675.

**14. Number of colours**

Number of colours used from the ten proposed colours. The ten proposed colours are black, grey, red, blue, dark green, light green, sky blue, brown, orange and yellow.

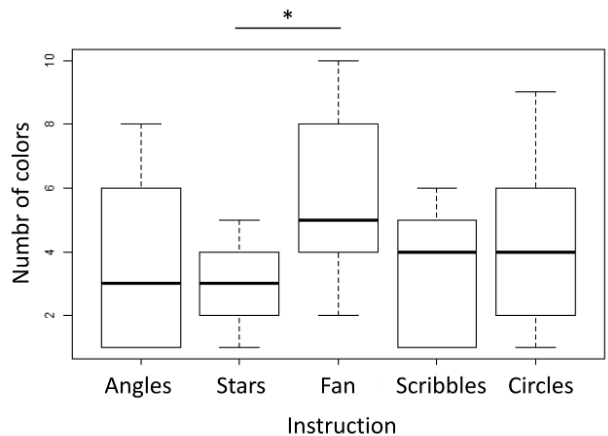

Figure S14: Boxplot of the number of colours according to instructions. Stars indicates significant difference between two instructions with p <0.05. Kruskal-Wallis test: p=0.04.

**Supplementary figures and tables referenced in the main text**

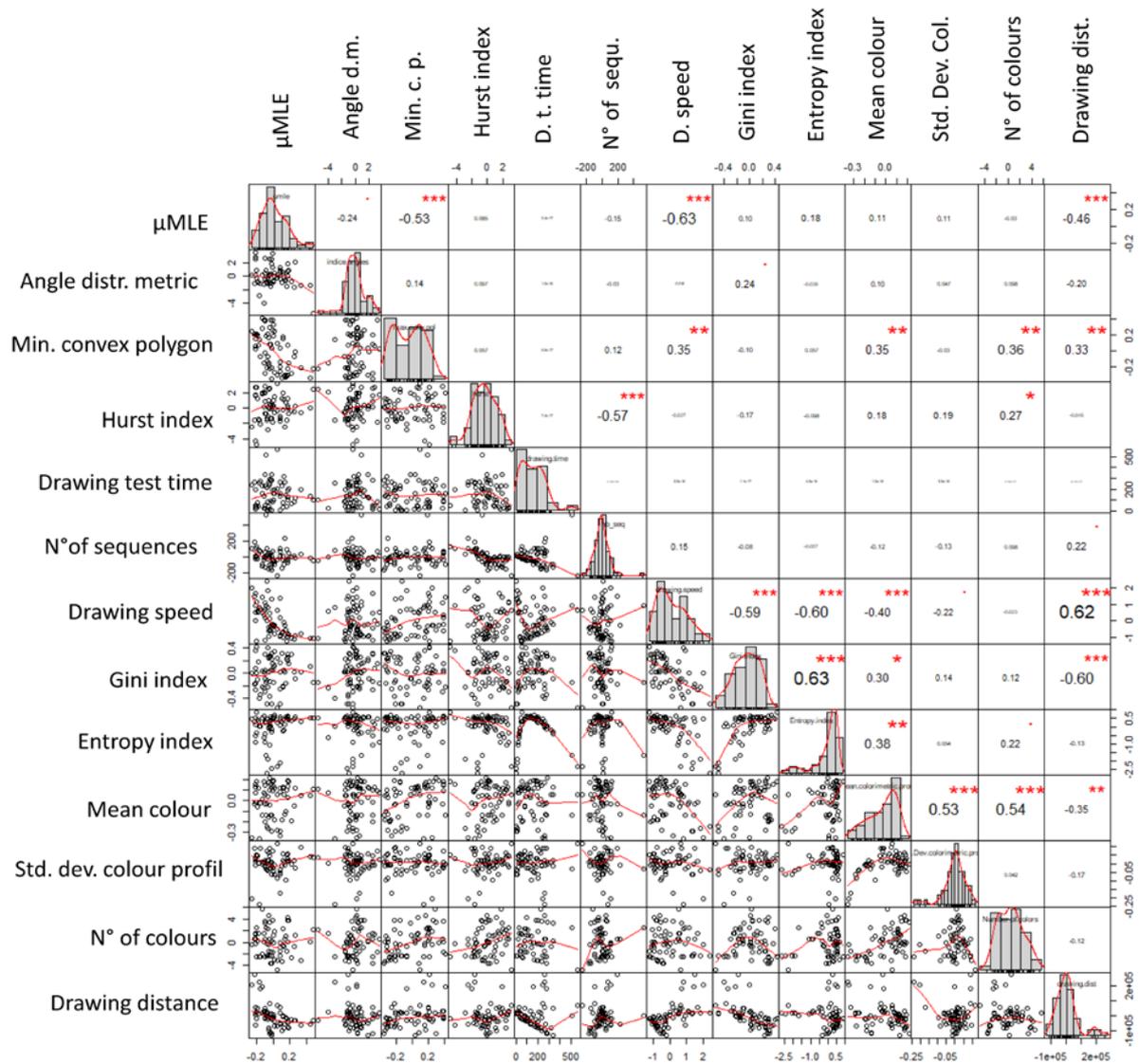

Figure S15: correlation chart of the 13 metrics for dataset#1 (first step, after removal of the drawing-time proportion and correction according to the drawing test time). The diagonal of the graph provides the distribution of each metric, whilst the bottom left and the top-right provide the correlation figure and the correlation coefficient between two metrics, respectively. Statistical value is given with the correlation coefficient: * means p <0.05, ** means p <0.01, and *** means p <0.001.

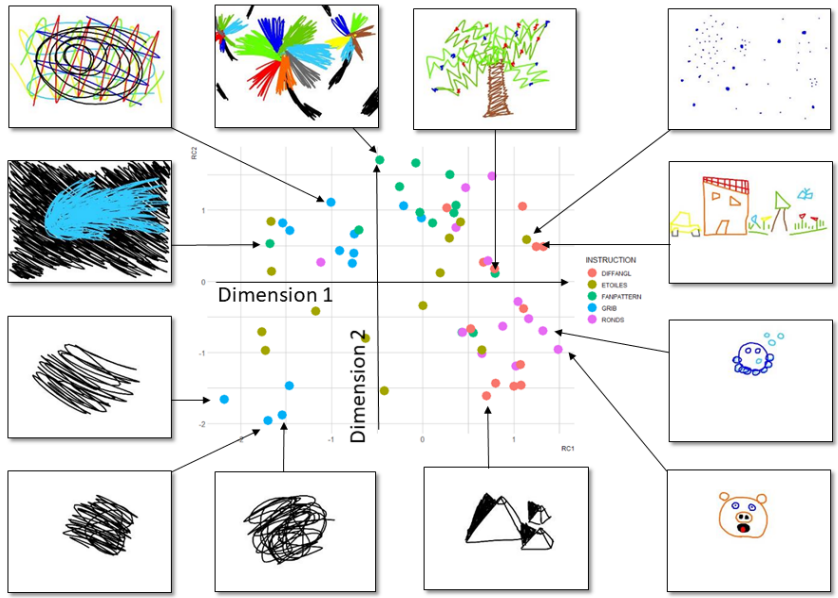

Figure S16a: Examples of Dataset#1 (first step) drawings according to Dimension 1 and Dimension 2, provided by the PCA. Dimension 1 may represent representativeness in drawing whilst Dimension 2 may represent diversity in drawings.

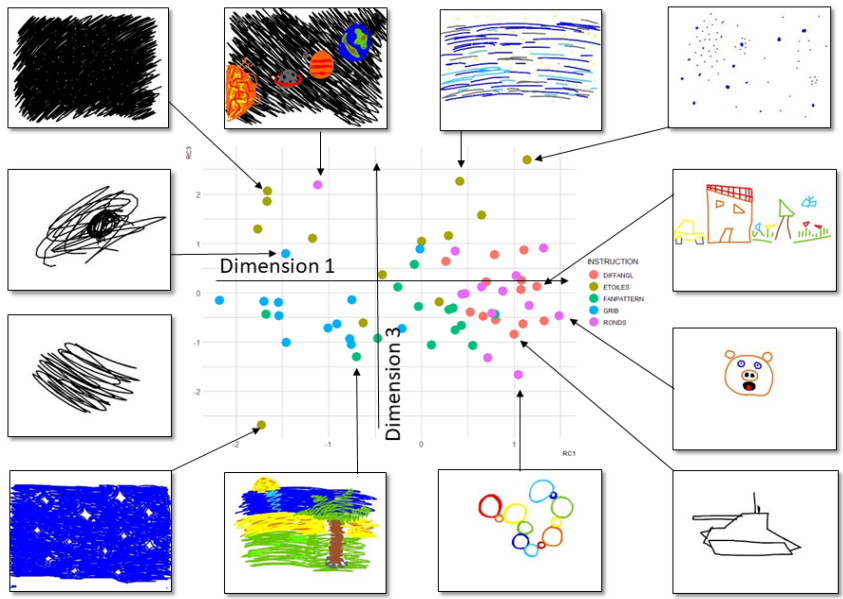

Figure S16b: Examples of Dataset#1 (first step) drawings according to Dimension 1 and Dimension 3, provided by the PCA. Dimension 1 may represent representativeness in drawing whilst Dimension 3 may represent periodicity in drawings.

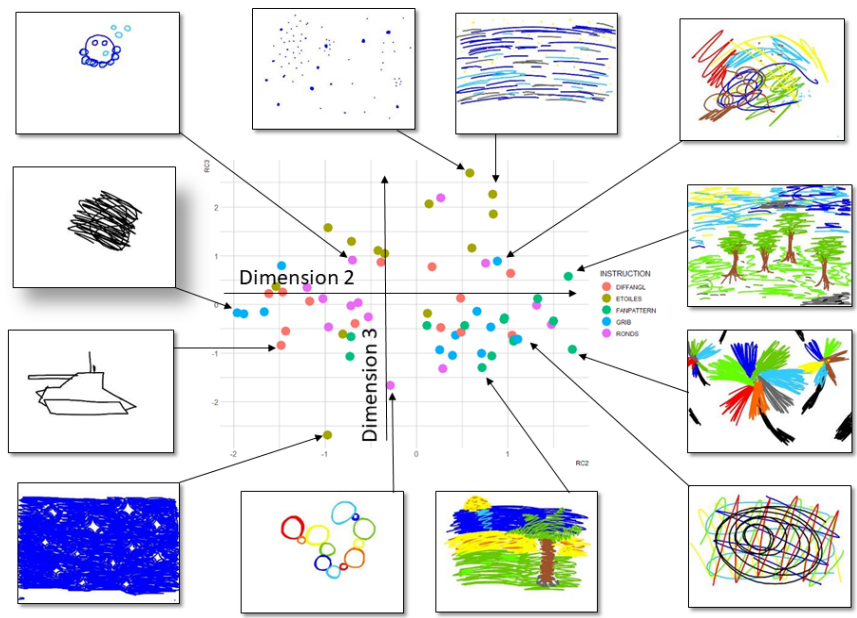

Figure S16c: Examples of Dataset#1 (first step) drawings according to Dimension 2 Dimension 3, provided by the PCA. Dimension 2 may represent diversity in drawing whilst Dimension 3 may represent periodicity in drawings.

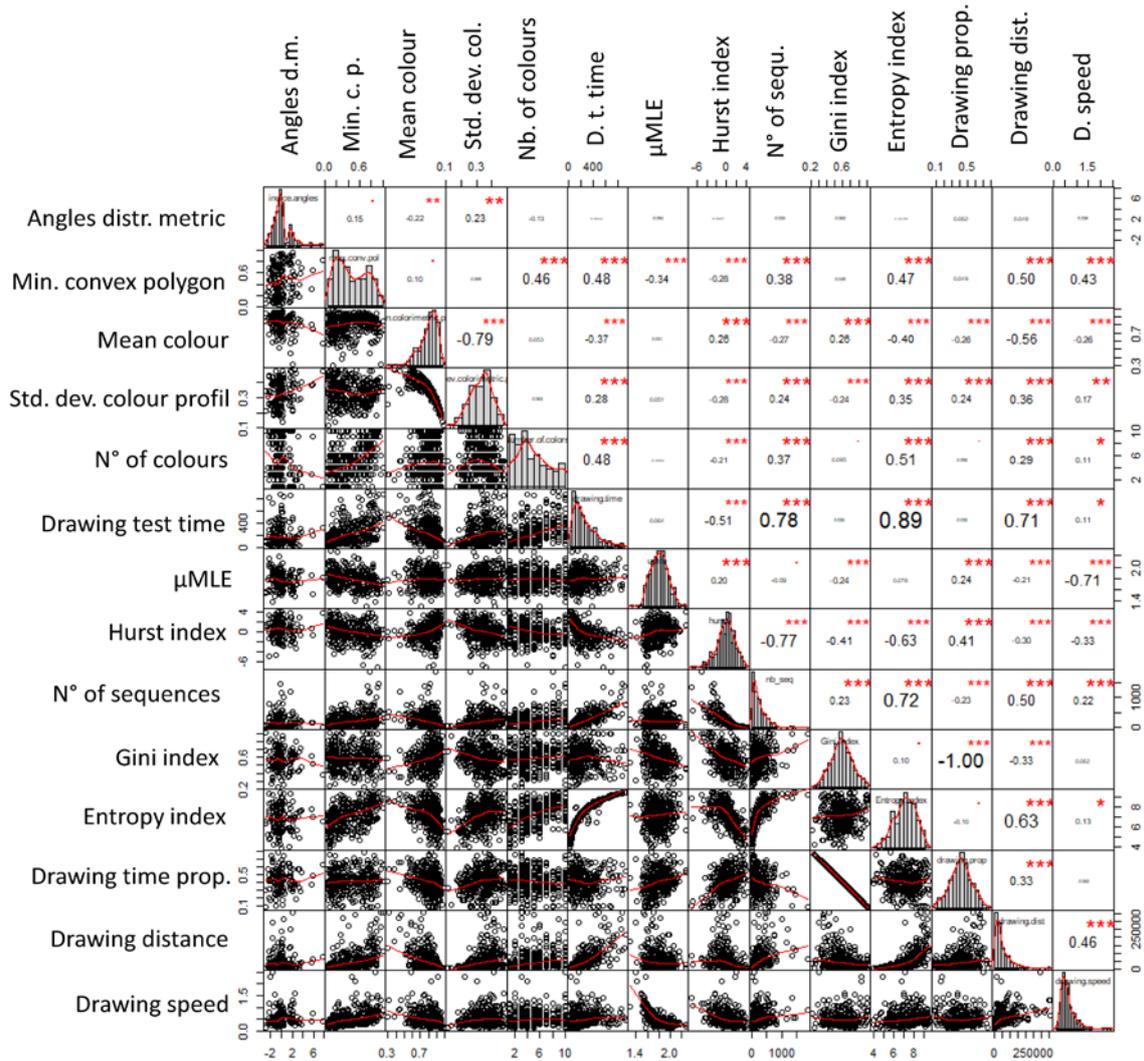

Figure S17: Correlation chart of the 14 metrics for dataset#2 (second step). The diagonal of the graph provides the distribution of each metric, whilst the bottom left and the top-right provide the correlation figure and the correlation coefficient between two metrics, respectively. Statistical value is given with the correlation coefficient: * means p <0.05, ** means p <0.01, and *** means p <0.001.

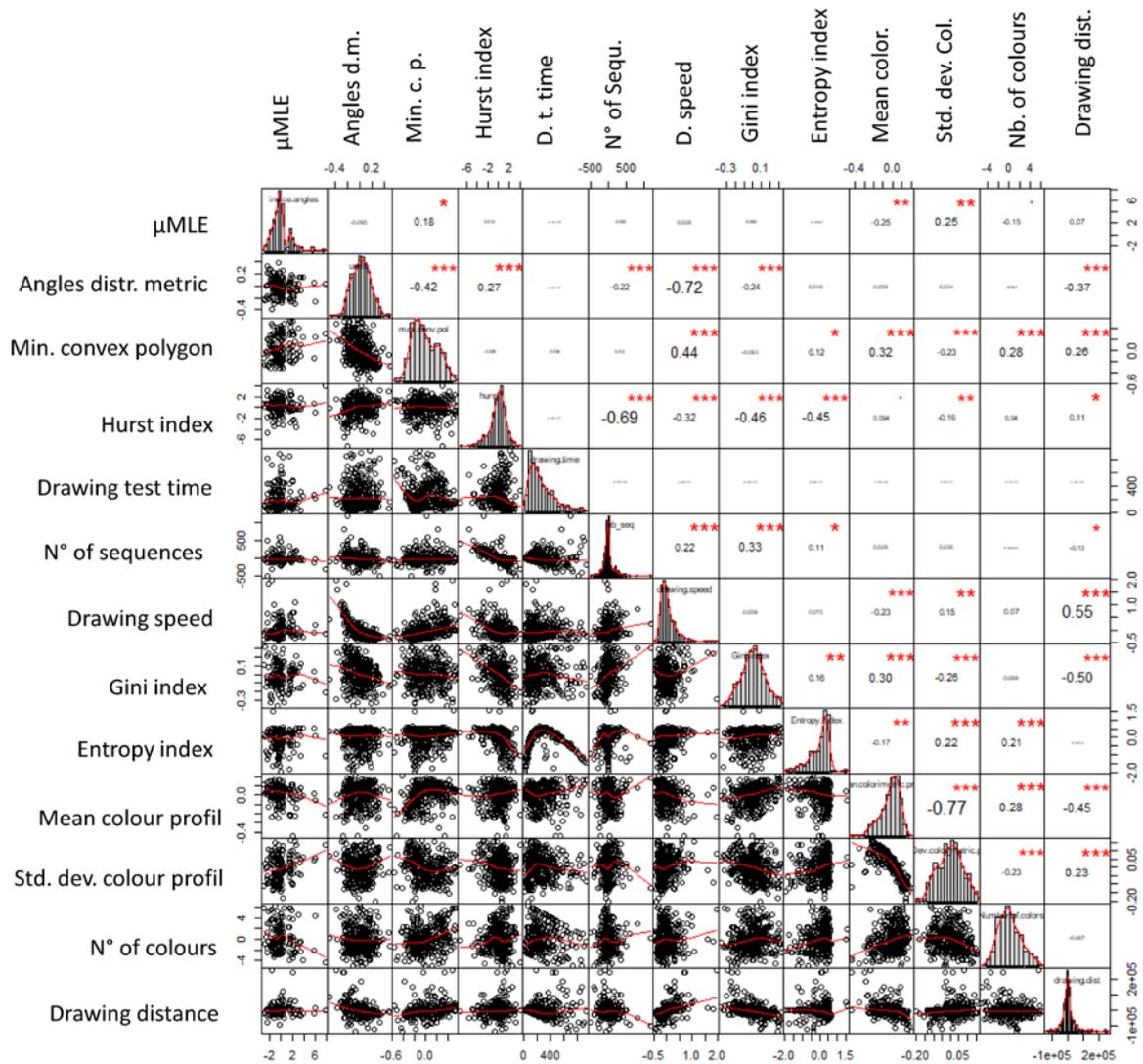

Figure S18: Correlation chart of the 13 metrics for dataset#2 (second step, after removal of the drawing-time proportion and correction according to the drawing test time). The diagonal of the graph provides the distribution of each metric, whilst the bottom left and the top-right provide the correlation figure and the correlation coefficient between two metrics, respectively. Statistical value is given with the correlation coefficient: * means p <0.05, ** means p <0.01, and *** means p <0.001.

Table S1: Loading of the 13 metrics on the three Varimax rotation PCA dimensions (before selection of variables with loading superior to 0.4) of Dataset#1

|  | Dim1 | Dim2 | Dim3 |
| --- | --- | --- | --- |
| µMLE | 0.639 | -0.373 | 0.314 |
| Angle distribution metric |  | 0.347 |  |
| Min. conv. pol. | -0.451 | 0.725 | -0.159 |
| Hurst index | -0.14 | 0.195 | 0.849 |
| Drawing test time |  |  |  |
| Number of sequences | -0.115 |  | -0.755 |
| Drawing speed | -0.911 |  | -0.12 |
| Gini index | 0.762 | 0.301 | -0.213 |
| Entropy index | 0.624 | 0.375 | -0.235 |
| Mean colour profile | 0.365 | 0.758 | 0.254 |
| Std. dev. colour profile | 0.221 | 0.296 | 0.398 |
| Number of colours |  | 0.716 |  |
| Drawing distance | -0.746 |  | -0.228 |

Table S2: Mann-Whitney test for each metric of each dimension between the scribbles of toddlers and those drawn by adults. p-values<0.05 are in bold print.

| Dimension | Metrics | W | p-value | Mean for toddlers | Mean for adults |
| --- | --- | --- | --- | --- | --- |
| 1 | µMLE | 101 | 0.225 | -0.07±0.14 | -0.12±0.09 |
|  | drawing distance | 47 | 0.097 | 30239±72983 | 26505±14882 |
|  | **drawing speed** | **29** | **0.007** | **0.31±0.73** | **1.04±0.68** |
| 2 | Mean colour Profil | 93 | 0.473 | 0.01±0.09 | -0.05±0.16 |
|  | **Number of colours** | **118** | **0.03** | **2.79±2.68** | **0.09±1.** |
| 3 | Hurst | 84 | 0.769 | 1.29±1.03 | 0.71±2.0 |
|  | **Number of sequences** | **11** | **<0.0001** | **-52.12±33.0** | **5.39±18.3** |